\documentstyle[prb,aps,multicol,float,epsfig,prb]{revtex}
\begin{document}
\draft
\title
{
Structural phase control of 
(La${}_{1.48}$Nd${}_{0.40}$Sr${}_{0.12}$)CuO${}_4$ thin films 
by epitaxial growth technique
}
\author{
I. Tsukada
}
\address{
Central Research Institute of Electric Power Industry, 
2-11-1 Iwado-kita, Komaeshi, Tokyo 201-8511, JAPAN 
}
\date{\today}
\maketitle

\begin{abstract}
Epitaxial growth of (La${}_{1.48}$Nd${}_{0.40}$Sr${}_{0.12}$)CuO${}_4$ 
thin films was studied by pulsed-laser deposition technique 
on three different substrates, SrTiO${}_3$ (100), LaSrAlO${}_4$ (001), 
and YAlO${}_3$ (001). 
The (Nd,Sr,Ce)${}_2$CuO${}_4$-type structure appears at the initial 
growth stage on SrTiO${}_3$ (100) when the film is deposited under the 
growth conditions optimized for (La,Sr)${}_2$CuO${}_4$. 
This (Nd,Sr,Ce)${}_2$CuO${}_4$-type structure can be eliminated 
by increasing the substrate temperature and the laser repetition 
frequency. Films on LaSrAlO${}_4$ (001) maintain 
a La${}_2$CuO${}_4$-type structure 
as bulk samples, but those on YAlO$_3$ (001) show phase separation 
into La${}_2$CuO${}_4$- and Nd${}_2$CuO${}_4$-type structures. 
Such complicated results are explained in terms of the competition 
between 
lattice misfit and thermodynamic conditions. 
Interestingly the films with La${}_2$CuO${}_4$-type structure prepared 
on 
SrTiO${}_3$ and LaSrAlO${}_4$ show different surface structures and 
transport properties. 
The results indicate the possibility of controlling charge stripes of 
(La${}_{1.48}$Nd${}_{0.40}$Sr${}_{0.12}$)CuO${}_4$ as was demonstrated 
in (La,Ba)${}_2$CuO${}_4$ thin films by Sato {\it et al.} 
(Phys. Rev. B {\bf 62}, R799 (2000)). 
\end{abstract}
\pacs{74.72.Jt, 74.76.Bz, 74.80.Dm}

\begin{multicols}{2}

\narrowtext
\section{Introduction}

One of the current issues of high-temperature (high-$T_c$) 
superconductivity is the formation of the stripe phase. 
Since the discovery of one-dimensional modulations of static spin 
and charge 
order in (La${}_{2-x-y}$Nd${}_y$Sr${}_x$)CuO${}_4$ (LNSCO) crystals,
\cite{Tranquada1} 
the relationship between the crystal structure and the stripe has 
been intensively studied.
\cite{Tranquada2} 
Thin-film samples have a great advantage in the study of this issue, 
because one can modify the lattice parameters from those 
of bulk samples through an epitaxial growth technique. 
Sato and Naito\cite{Sato1} and Locquet {\it et al.}\cite{Locquet1} 
have shown that in-plane compression of the crystal significantly 
increases $T_c$ of (La${}_{2-x}$Sr${}_x$)CuO${}_4$ (LSCO) films grown 
on a LaSrAlO${}_4$ (001) substrate. 
More surprisingly, the films on LaSrAlO${}_4$ (001) do not exhibit the 
anomalous suppression of $T_c$ at a carrier concentration of $x \approx$ 
0.12 which has been studied for a decade as a 1/8 problem. 
The recovery of superconductivity has already been confirmed in LSCO and 
(La${}_{2-x}$Ba${}_x$)CuO${}_4$ films. 
\cite{Sato2,Sato3} 
Nevertheless, it is important to study this issue using LNSCO thin films, 
because LNSCO is the only composition in which the static charge order 
has been confirmed experimentally.
\cite{Fujita1} 
In this paper, I report the detailed growth conditions of LNSCO thin films 
on several substrates. 
In contrast to the case of LSCO and LBCO, the presence of Nd ions 
requires more careful adjustment of growth conditions, because one 
must pay attention to three different phases.

\section{Sample Preparation}

All the films were prepared by pulsed laser deposition (PLD). 
A KrF ($\lambda$ = 248~nm) excimer laser was used with an energy 
density of approximately 1.5~J/cm${}^2$. 
In order to optimize the growth condition, I focused on the substrate 
temperature ($T_s$) and the growth rate, which is controlled by 
changing the laser repetition rate ($f$). 
A typical growth rate per pulse is about 0.5~{\AA}. 
Because of the absence of any report on LNSCO-film growth, 
the reported growth conditions of LSCO films were used as the the 
starting point for this study. 
According to the previous literature
\cite{Trofimov1,Si1} 
typical values of $T_s$ for LSCO films are around 750 $\sim$ 
780${}^{\circ}$C. 
As will be shown later, however, LNSCO film requires a higher substrate 
temperature than LSCO films. 
A target with a chemical composition of La:Nd:Sr:Cu = 
1.48:0.40:0.12:1.00, which is the same composition 
as that reported in Ref.~[1], was prepared by a conventional 
solid-state reaction method. 
Oxidation was carried out with a mixture gas of O${}_2$ and O${}_3$; 
total pressure ($P_{{\rm O}_2 + {\rm O}_3}$) was held at approximately 
3.9~Pa (30~mTorr) and the partial pressure of O${}_3$ was held at 
approximately 5.0 $\times$ 10${}^{-3}$~Pa during the deposition. 
After the deposition, total pressure was increased to 26.3~Pa (200~mTorr) 
without changing the partial pressure of O${}_3$, and films were annealed 
under this condition for 30 min. Oxidation was terminated 
when the films had cooled to 200${}^{\circ}$C. 
All the growth conditions and grown phases are summarized in 
Table~\ref{GrowthCondition1}, 
in which the main and secondary phases are shown.

\section{Results and Discussion}

\subsection{$T$, $T^*$, and $T$' phases}

\begin{figure}
\includegraphics*[width=80truemm]{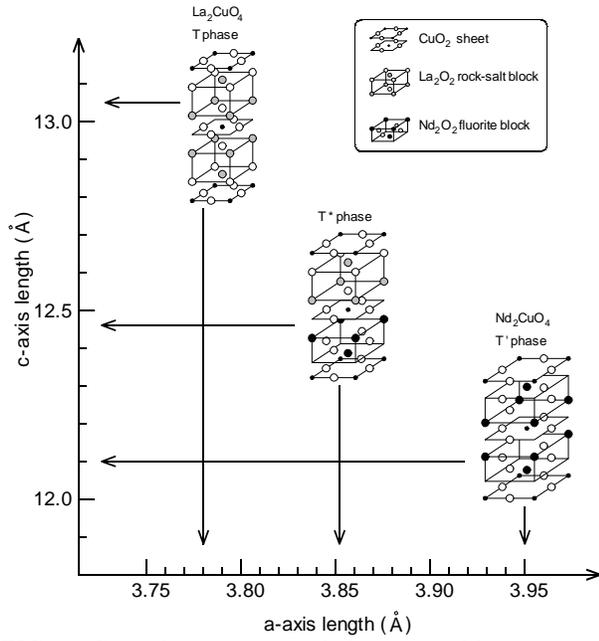}
\caption{Crystal structures and their typical lattice parameters for 
$Ln_2$CuO${}_4$-type compounds. 
T-phase structure includes rock-salt-type 
La${}_2$O${}_2$ block between neighboring CuO${}_2$ sheets, 
while T'-phase structure includes fluorite Nd${}_2$O${}_2$ block 
between them. 
T${}^*$ includes both the rock-salt and fluorite blocks, 
which appear alternately. 
In going from $T$ to $T'$, the $a$-axis length increases 
and the $c$-axis length decreases.}
\label{Fig.1}
\end{figure}

It is useful to describe the crystal structures before 
showing the experimental results. 
Figure~\ref{Fig.1} shows the schematic crystal structure of three 
$Ln_2$CuO${}_4$-type ($Ln$: lanthanoid) compounds: 
$T$, $T^*$, and $T$' phases. 
The rock-salt (La${}_2$O${}_2$) and fluorite (Nd${}_2$O${}_2$) 
blocks are inserted between adjacent CuO${}_2$ sheets, 
in the case of $T$ and $T$' phases, respectively.
\cite{Tetra1} 
Singh {\it et al.} reported that Nd substitution for La in La${}_2$CuO${}_4$ 
causes a phase transition from the $T$ to the $T$' phase in accordance 
with the continuous decrease of ionic radius of the La-site element.
\cite{Singh1} 
For the $T^*$ phase, these two blocks are alternately inserted 
between the adjacent CuO${}_2$ planes. 
In a bulk sample, it is believed to be necessary to perform complicated 
substitution to obtain the $T^*$-phase structure. 
For example, in the first $T^*$-phase superconductor 
(Nd,Sr,Ce)${}_2$CuO${}_4$, 
Sr${}^{2+}$ ions in the rock-salt block are partially substituted by 
Nd${}^{3+}$ ions, and simultaneously, Nd${}^{3+}$ ions in the fluorite 
block are partially substituted by Sr${}^{2+}$ and Ce${}^{4+}$ ions.
\cite{Sawa1}

Because of the difference in the intrinsic size of rock-salt and fluorite 
blocks, the lattice parameters of the three phases are clearly 
different from one another. As is indicated by arrows in Fig.~\ref{Fig.1},
\cite{Tetra1} 
the representative lattice parameters at room temperature are 
$a_0$ = 3.80~{\AA}, $c_0$ = 13.1~{\AA},
\cite{Goto1} 
$a_0$ = 3.85~{\AA}, $c_0$ = 12.6~{\AA},
\cite{Sawa1,Kwei1,Izumi1} 
and $a_0$ = 3.95~{\AA}, $c_0$ = 12.1~{\AA}
\cite{Singh1,Tokura1} 
for $T$, $T^*$, and $T$' phases, respectively.
One might expect the boundaries from $T$ to $T^*$ and $T^*$ to $T$' phases 
to become unclear after complicated chemical substitutions, 
however, in x-ray diffraction, with respect to the 00$l$ reflections, 
we can easily detect the $T^*$ phase by observing 00$l$ ($l$ : odd) 
reflections which are absent in both $T$- and $T$'-phase structures.

\subsection{Films on SrTiO${}_3$ (100)}

\begin{figure}
\includegraphics*{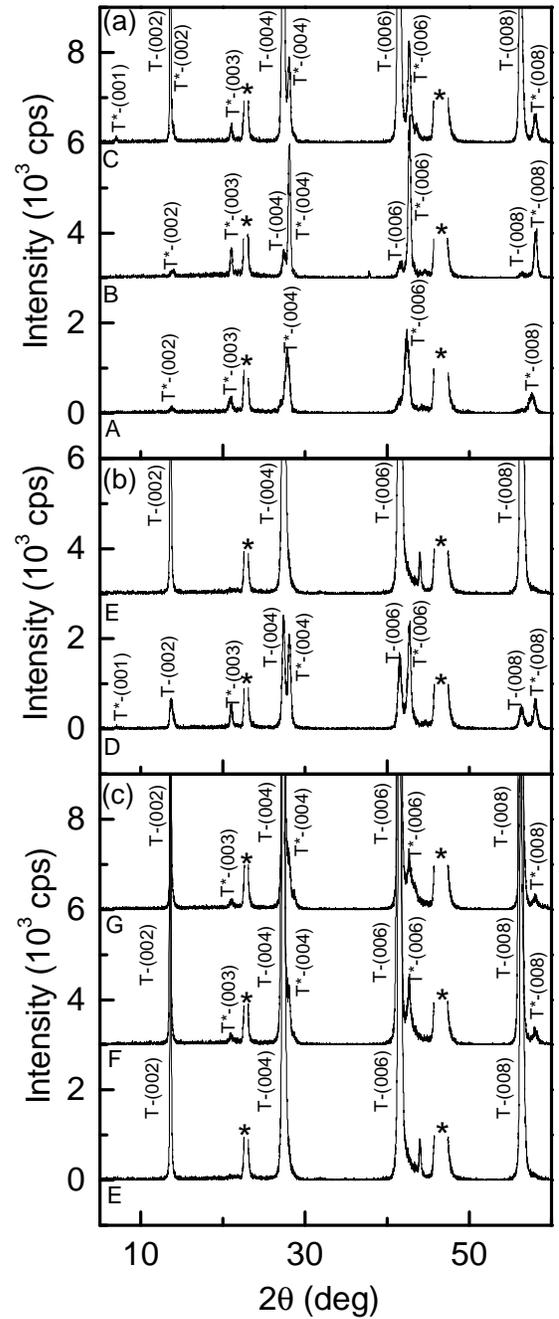}
\caption{X-ray diffraction spectra of the films 
on SrTiO${}_3$ (100): 
(a) deposition-time dependence at $f$ = 1~Hz, 
(b) deposition-time dependence at $f$ = 5~Hz, and 
(c) substrate-temperature dependence at $f$ = 5~Hz. 
$*$ indicates the peaks attributed to the substrates.}
\label{Fig.2}
\end{figure}

First, I show the results of films grown on SrTiO${}_3$ (100) 
($a_0$ = 3.905~{\AA}). 
Figure~\ref{Fig.2} (a) shows the x-ray diffraction spectra for samples 
A ($t$ = 150~{\AA}), B ($t$ = 300~{\AA}), and C ($t$ = 1500~{\AA}) 
prepared at $f$ = 1~Hz. 
For sample A, almost all observed peaks are assigned to a $T^*$-phase 
structure, with only a trace of the $T$ phase being observable. 
By increasing the thickness, however, the diffraction peaks of the 
$T$ phase increase in sample B, and the volume fraction of $T$ phase 
exceeds that of the $T^*$ phase in sample C. 
This result indicates that the $T^*$ phase is stably grown only at 
the initial growth stage. 
The $T$ phase appears after the film thickness exceeds a certain value. 
A similar tendency was observed for samples D ($t$ = 300~{\AA}) and E 
($t$ = 1500~{\AA}) (Fig.~\ref{Fig.2} (b)), 
which were prepared at a faster pulse rate, $f$ = 5~Hz. 
Interestingly, the $T^*$ phase is no longer a major phase in sample D 
and almost disappears in sample E. 
When we compare samples C and E (and also samples B and D), 
it is clear that the faster growth rate suppresses 
the appearance of the $T^*$ phase. 
Since the film thickness is almost the same between samples C and E 
(and between samples B and D), the diffraction data actually reflect 
the difference in the volume fractions. 
Thus the laser repetition rate is concluded to be one of 
the crucial factors in obtaining a $T$-phase film.

Next let us see the $T_s$ dependence at $f$ = 5~Hz. 
X-ray diffraction spectra of three samples prepared at 
$T_s$ = 830${}^{\circ}$C (sample E), 
800${}^{\circ}$C (sample F), and 750${}^{\circ}$C (sample G) 
are shown in Fig.~\ref{Fig.2} (c). 
The $T^*$ phase is again found in samples F and G. 
In contrast to the previous reports on the PLD growth of LSCO thin films, 
in which the optimum substrate temperatures were reported to be 
750 $\sim$ 780${}^{\circ}$C,
\cite{Trofimov1,Si1}
the $T^*$ phase remains in LNSCO films even at $T_s$ = 800${}^{\circ}$C. 
Therefore, the Nd substitution probably creates a stable $T^*$-phase region, 
and pushes the stable $T$-phase region to higher temperatures.

As a result, it is found that $T$-phase LNSCO films can be grown 
on SrTiO${}_3$ (100) only at $T_s$ = 830${}^{\circ}$C and $f$ = 5~Hz, 
which are higher and faster than the values for LSCO films. 
The difference between samples A, B and C is explained as follows. 
The longer lattice constant of SrTiO${}_3$ first stabilizes 
$T^*$-phase LNSCO, but, after sufficient lattice relaxation, 
$T$-phase LNSCO becomes stable. 
This means that all treatments which {\em suppress} the epitaxial effect 
from SrTiO${}_3$ will stabilize the $T$-phase structure. 
The present result indicates that the faster growth rate reduces the 
influence of lattice mismatch with the substrate.

\subsection{Films on LaSrAlO${}_4$ (001) and YAlO${}_3$ (001)}

\begin{figure}
\includegraphics*{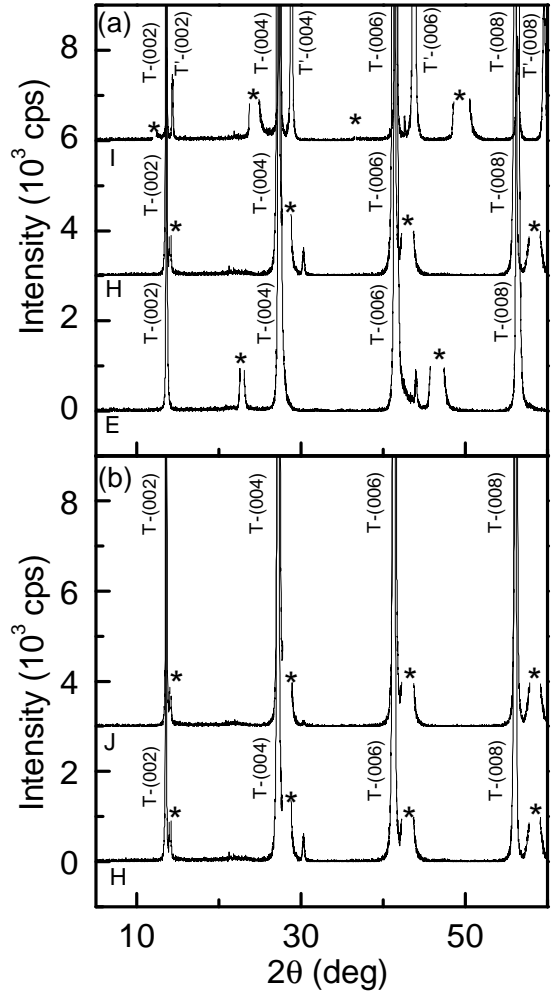}
\caption{(a) X-ray diffraction sprectra of the films on different 
substrates with $T_s$ = 830${}^{\circ}$C and $f$ = 5~Hz. 
(b) X-ray diffraction spectra of the films on LaSrAlO${}_4$ (001) 
with different laser repetition rates. 
$*$ indicates the peaks attributed to the substrates.}
\label{Fig.3}
\end{figure}

To check the effect of lattice matching, films on LaSrAlO${}_4$ (001) 
($a_0$ = 3.76~{\AA}) and YAlO${}_3$ (001) ($a_0$ = 3.71~{\AA}) 
were also prepared.
\cite{YAP1}
Growth conditions were set at $T_s$ = 830${}^{\circ}$C and $f$ = 5~Hz 
similarly to the case of pure $T$-phase films on SrTiO${}_3$. 
Results are shown in Fig.~\ref{Fig.3} (a). 
First, one can see that sample H consists of a pure $T$-phase structure, 
as does sample E. The $c$-axis length of sample H is approximately 
13.13~{\AA}, which is significantly longer than that of 
sample E, 13.07~{\AA}. 
This $c$-axis expansion is understood to be a consequence of in-plane 
compression by LaSrAlO${}_4$ (001) as was reported for LSCO films.
\cite{Sato1,Locquet1,Sato2}
A substantial difference from the case of SrTiO${}_3$ is 
that the $T^*$ phase is not stable on LaSrAlO${}_4$ 
even if we reduce the growth rate. 
As is shown in Fig.~\ref{Fig.3} (b), the film prepared at $f$ = 1~Hz 
also consists of a pure $T$-phase structure, indicating 
that the appearance of the $T^*$ phase is a consequence of only 
the in-plane lattice expansion by SrTiO${}_3$.

The film on YAlO${}_3$ (001) shows a complex structure. 
Sample I shows not only a set of Bragg peaks from the $T$ phase but 
also another set of peaks at higher angles with similar intensities. 
It is difficult to identify this second phase from only the $c$-axis 
length, because the estimated $c$-axis length is 12.41~{\AA}, 
which is an intermediate value between those of $T^*$ and $T'$ phases. 
However, the absence of (00$l$) ($l$ : odd) peaks indicates 
that the second phase has a $T'$-phase structure. 
Recently reported (La,Ce)${}_2$CuO${}_4$ thin films grown 
on SrTiO${}_3$ (100) also have similar $c$-axis lengths.
\cite{Matsuo1,Naito1} 
The reason why the $T'$ phase was stably grown on YAlO${}_3$ (001) 
is not simple, because the in-plane lattice parameters of 
YAlO${}_3$ (001) are very short (shorter than those of the 
$T$ phase), while the $a$-axis length of a typical $T'$-phase 
crystal is far longer than those of $T$ and $T^*$ phases. 
Therefore, the lattice misfit between YAlO${}_3$ and the $T'$ phase 
takes the largest value.

\subsection{Phase diagram}

\begin{figure}
\includegraphics*[width=80mm]{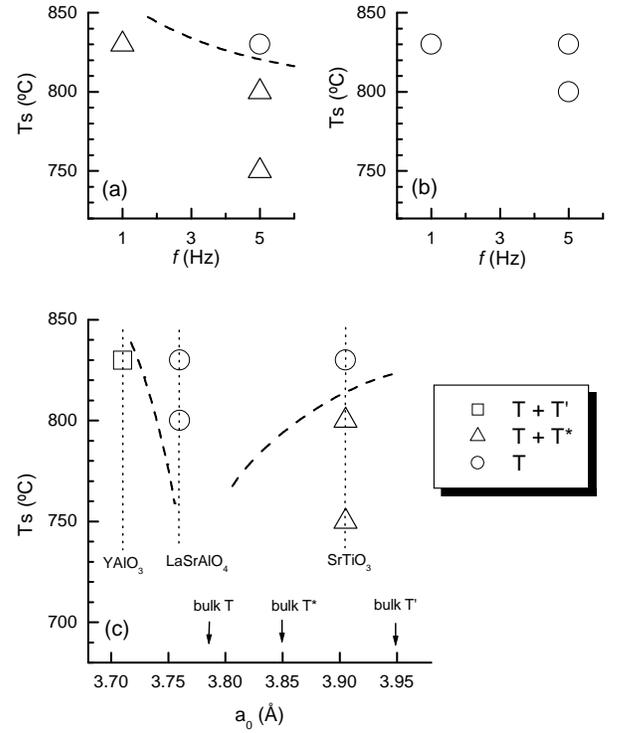}
\caption{A phase diagram of the grown phases as functions of 
$T_s$, $f$, and $a$-axis lattice constant of the substrates. 
Open circles, open triangles, and open squares shows pure $T$ phase, 
$T$ + $T^*$ phases, and $T$ + $T$' phases, respectively. 
(a) $T_s$ and $f$ dependences of the grown phase on SrTiO${}_3$ (100). 
(b) $T_s$ and $f$ dependences of the grown phase on LaSrAlO${}_4$ (100). 
(c) $T_s$ and $a_0$ dependences of the grown phase on three substrates. 
Dotted lines indicate the $a$-axis lengths of three substrates. 
Dashed lines are guides to the eye and roughly indicates 
the stable $T$-phase region. 
}
\label{Fig.4}
\end{figure}

Based on the x-ray diffraction results, I can summarize the obtained 
phase of (La${}_{1.48}$Nd${}_{0.40}$Sr${}_{0.12}$)CuO${}_4$ films 
as a function of $T_s$, $f$, and the $a$-axis length of substrates. 
Figure~\ref{Fig.4} (a) clearly indicates that higher growth temperatures 
and faster growth rates are indispensable for eliminating the $T^*$ phase 
on SrTiO${}_3$ (100). 
There is no such difficulty in obtaining the $T$-phase structure 
in the films on LaSrAlO${}_4$, as shown in Fig.~\ref{Fig.4} (b). 
To elucidate the difference between SrTiO${}_3$ and LaSrAlO${}_4$, 
it is useful to plot the phase diagram as a function of the $a$-axis 
length. Figure~\ref{Fig.4} (c) summarizes the obtained phases 
at $f$ = 5~Hz. It is clear that the $T$-phase region extends 
to low temperatures at around $a_0$ = 3.80~{\AA}. 
Since LaSrAlO${}_4$ shows the least lattice misfit with the bulk 
(La${}_{1.48}$Nd${}_{0.40}$Sr${}_{0.12}$)CuO${}_4$, the extension of the 
$T$-phases region to lower temperatures is reasonable.

The $a$-axis length of YAlO${}_3$ is far shorter than that of bulk 
(La${}_{1.48}$Nd${}_{0.40}$Sr${}_{0.12}$)CuO${}_4$; the difference of 
almost 0.1~{\AA} leads to the large misfit value of 2.4{\%}. 
In the case of such a lattice-mismatch condition, we may expect 
several situations. One is that the phase is governed only by 
thermodynamic conditions with no influence of the substrates. 
However, this does not seem to be the current case 
because the thermodynamic conditions favor only one stable phase 
and the coexistence of $T$ and $T$' phases cannot be explained. 
Another possibility is that the short $a$-axis length actually 
exerts a considerable strain effect on the film. 
In some regions, the very short $a_0$ of YAlO${}_3$ allows Nd-poor films 
to be grown epitaxially, which leads to the growth of Nd-rich films 
in other regions which are less influenced by the substrate. 
This scenario is more likely for the current case, because it naturally 
leads to the phase separation of $T$ and $T$' phases.

The appearance of the $T^*$ phase in the film on SrTiO${}_3$ is 
unexpected and suggests a strong epitaxial effect. 
The $a$-axis length of SrTiO${}_3$ is far longer 
than that of bulk (La${}_{1.48}$Nd${}_{0.40}$Sr${}_{0.12}$)CuO${}_4$. 
As is indicated in Fig.~\ref{Fig.4}, the typical $a$-axis length of 
the bulk $T^*$-phase structure is between those of the bulk $T$ 
phase and SrTiO${}_3$. 
Therefore it is most likely that the competition between the 
thermodynamic conditions and the epitaxial effect forces LNSCO films 
to have a $T^*$-phase structure. 
Note that the $T^*$ phase has not previously been identified in the 
LNSCO compound; 
the Nd substitution for La in LSCO induces a direct phase 
transition from the $T$ to the $T$' phase.
\cite{Singh1} 
Thus it is concluded that the appearance of the $T^*$ phase is 
the result of an epitaxial strain effect. 
The absence of $T$' phase in this case suggests that the film on 
SrTiO${}_3$ is homogeneous, in contrast to that on YAlO${}_3$. 
The present result indicates that moderate expansion along the in-plane 
direction can stabilize the $T^*$-phase structure of LNSCO. 
One may ask how rock-salt and fluorite blocks are 
formed from the same La-Nd-Sr components. 
There is no direct evidence, but I suggest that the rock-salt block may be 
slightly La rich while the fluorite block may be Nd rich.

In closing this section, the effects of the lattice expansion 
coefficient should be discussed. The thermal expansion coefficients of 
the three substrate materials are reported to be 
$\alpha$ = 9 $\times$ 10${}^{-6}$~K${}^{-1}$, 
10.5 $\times$ 10${}^{-6}$~K${}^{-1}$, 
and 5-10 $\times$ 10${}^{-6}$~K${}^{-1}$, 
for SrTiO${}_3$ (100), LaSrAlO${}_4$ 
(001), and YAlO${}_3$ (001), respectively.
\cite{Locquet1,Asano1} 
Expected in-plane lattice constants at the optimal growth temperature 
(830${}^{\circ}$C) are then $a$ = 3.914~{\AA}, 3.77~{\AA}, 
and 3.71-3.72~{\AA}. 
Since, to our knowledge, there is no report on the high-temperature 
thermal expansion coefficient of LNSCO compounds, we refer to 
the values for LSCO. 
Using the value reported by Locquet {\it et al.} 
($\alpha$ = 8.5 $\times$ 10${}^{-6}$~K${}^{-1}$)
\cite{Locquet1}, 
we can expect $a$ = 3.81~{\AA} for the present composition of LNSCO. 
The shift of lattice constants is very small, and the epitaxial 
relationship at the growth temperature is similar to that 
at room temperature. 
Therefore, we may safely discuss the epitaxial relationship using 
the room-temperature lattice constants.

\subsection{Comparison of $T$-phase films on SrTiO${}_3$ and LaSrAlO${}_4$}

\begin{figure}
\includegraphics*{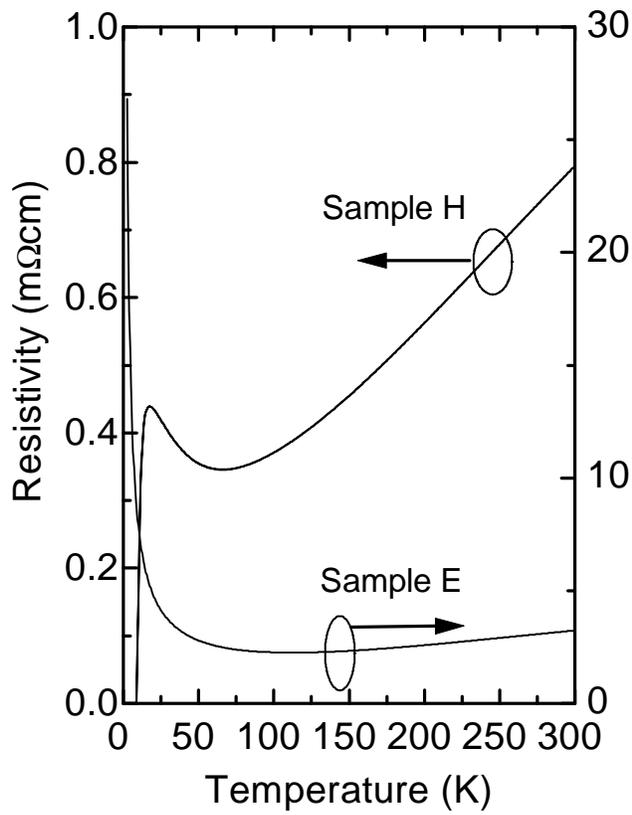}
\caption{Temperature dependence of the resistivity of samples E and H, 
on SrTiO${}_3$ (100) and LaSrAlO${}_4$ (001), respectively. 
Sample H shows superconductivity below, while sample E becomes insulating 
toward $T$ = 0~K.}
\label{Fig.5}
\end{figure}

In this section, I address several properties of the pure 
$T$-phase films on SrTiO${}_3$ (100) (Sample E) and LaSrAlO${}_4$ (001) 
(Sample H), because they show a clear difference. 
One of the important differences is in the transport property. 
Figure~\ref{Fig.5} shows the temperature dependence of the resistivity 
for samples E and H. 
Neither sample exhibits the discontinuous jump associated with the 
the structural phase transition from the low-temperature orthorhombic 
(LTO) to the low-temperature tetragonal (LTT) phase, 
which is observed in bulk LNSCO crystals. 
This indicates that a strong epitaxial relation prevents the films 
from behaving as bulk.
\cite{Tetra2} 
Sample H shows sufficiently low resistivity at room temperature 
and becomes superconducting below $T_c$ = 8.5~K, 
while sample E has a high resistivity and shows insulating behavior. 
Nakamura and Uchida presented the resistivity and magnetization data 
of (La${}_{1.48}$Nd${}_{0.40}$Sr${}_{0.12}$)CuO${}_4$ single crystals; 
the former indicates superconductivity below 10~K 
while the latter indicates almost no diamagnetism.
\cite{Nakamura1} 
Subsequent to several investigations on the relationship 
between the crystal structure and superconductivity, 
\cite{Crawford1,Buchner1} 
Moodenbaugh {\it et al.,} studied polycrystalline LNSCO in detail, 
upon which they found an absence of bulk superconductivity in the sample 
with the same composition as used in the current study.
\cite{Moodenbaugh1} 
The absence of superconductivity with the current chemical composition is 
still an open question, because it strongly depends on the preparation 
technique. 
However, at least the suppression of superconductivity in this 
composition is widely accepted, and the difference 
between samples E and H is substantial. 
The result is qualitatively similar to that reported in LSCO and LBCO 
thin films,
\cite{Sato2,Sato3} 
which suggests that the same epitaxial strain effect controls the 
superconductivity of LNSCO films. 
However, the high resistivity of sample E cannot be explained by only a 
strain effect, and I believe that significant lattice expansion 
due to the effect of the SrTiO${}_3$ substrate may induce a structural 
disorder that causes a very high resistivity value.

\begin{figure}
\includegraphics*{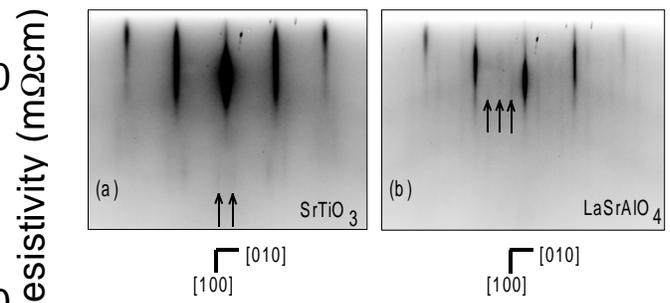}
\caption{RHEED observation of samples E and H. 
Arrows indicate the positions of satellite streaks.}
\label{Fig.6}
\end{figure}

The difference of the substrate also induces different 
structural modulations. 
Reflection high-energy electron diffraction (RHEED) patterns 
of samples E and H are shown in Fig.~\ref{Fig.6}. 
Several satellite streaks (indicated by arrows) are observed, 
in addition to the main streaks, for both films, 
but the positions are different from each other. 
In sample E, the structural modulation runs along the Cu-Cu diagonal 
direction with respect to the CuO${}_2$ plane, 
while it runs along the Cu-O-Cu bond direction in sample H. 
The difference in the modulation directions is probably related to the 
crystal symmetry of samples E and H. At present, I have no evidence that 
the observed modulations are related directly to the transport properties; 
the on-going study of the structural analysis will clarify this.

\section{Summary}

I have found that the growth conditions of $T$-phase LNSCO films 
are not the same as those of LSCO, and we should pay special attention 
not only to the substrate conditions (materials and temperatures) 
but also to the deposition rate. 
However, pure $T$-phase LNSCO films can be grown on SrTiO${}_3$, 
which opens the way to the study of the stripe problem with thin-film 
samples. 
The next step is to elucidate how the structural difference of LNSCO films 
stabilizes the static stripe order, by various methods. 
For this purpose, a detailed study of the dependence of structural 
and transport properties on Sr doping is now being performed, 
which will be reported elsewhere.

The author acknowledges H. Sato, M. Naito, Y. Ando, S. Komiya, and 
A. N. Lavrov for fruitful discussions.

\begin{table}
\caption{Growth conditions of the prepared films. }
\begin{tabular}{ccccccc}
Sample & Substrate & frequency & $T_s$ & thickness & \multispan{2}{\hfil phase\hfil} \\
       &           & [Hz]      & [C${}^{\circ}$] & [{\AA}] & main & sub \\
\hline
A & SrTiO${}_3$ (100)   & 1 & 830 &  150 & $T^*$ & $T$   \\
B & SrTiO${}_3$ (100)   & 1 & 830 &  300 & $T^*$ & $T$   \\
C & SrTiO${}_3$ (100)   & 1 & 830 & 1500 & $T$   & $T^*$ \\
D & SrTiO${}_3$ (100)   & 5 & 830 & 300 & $T$   & $T^*$ \\
E & SrTiO${}_3$ (100)   & 5 & 830 & 1500 & $T$   & ---   \\
F & SrTiO${}_3$ (100)   & 5 & 800 & 1500 & $T$   & $T^*$ \\
G & SrTiO${}_3$ (100)   & 5 & 750 & 1500 & $T$   & $T^*$ \\
H & LaSrAlO${}_4$ (001) & 5 & 830 & 1500 & $T$   & ---   \\
I & YAlO${}_3$ (001)    & 5 & 830 & 1500 & $T$  & $T$'   \\
J & LaSrAlO${}_4$ (001) & 1 & 830 & 1500 & $T$   & ---   \\
\end{tabular}
\label{GrowthCondition1}
\end{table}

\end{multicols}


\begin{references}
\bibitem{YAP1} We estimate the lattice parameters of YAlO${}_3$ using 
pseudotetragonal notation. To say exactly, YAlO${}_3$ is orthorhombic 
with $a_0$ = 5.18~{\AA} and $b_0$ = 5.33~{\AA}.
\bibitem{Tetra2} Preliminary structural anlysis indicates that both 
samples E and H have tetragonal symmetry at room temperature.
\end{references}
\end{document}